\documentclass[aps,prl,twocolumn,superscriptaddress,floatfix]{revtex4}

\usepackage{graphicx}
\usepackage{epstopdf}
\usepackage{times}

\begin{document}

\title{Velocity enhancement by synchronization of magnetic domain walls}

\author{Ale\v{s} Hrabec}
    \altaffiliation{These authors contributed equally to the work}
    \affiliation{Laboratoire de Physique des Solides, CNRS, Universit\'{e} Paris-Sud, Universit\'{e} Paris-Saclay, 91405 Orsay Cedex, France}
\author{Viola K\v{r}i\v{z}\'{a}kov\'{a}}
    \altaffiliation{These authors contributed equally to the work}
    \affiliation{Institut N\'{e}el, CNRS, 25 avenue des Martyrs, B.P. 166, 38042 Grenoble Cedex 9, France}
\author{Stefania Pizzini}
    \affiliation{Institut N\'{e}el, CNRS, 25 avenue des Martyrs, B.P. 166, 38042 Grenoble Cedex 9, France}
\author{Jo\~{a}o Sampaio}
    \affiliation{Laboratoire de Physique des Solides, CNRS, Universit\'{e} Paris-Sud, Universit\'{e} Paris-Saclay, 91405 Orsay Cedex, France}
\author{Andr\'{e} Thiaville}
    \affiliation{Laboratoire de Physique des Solides, CNRS, Universit\'{e} Paris-Sud, Universit\'{e} Paris-Saclay, 91405 Orsay Cedex, France}
\author{Stanislas Rohart}
    \email{stanislas.rohart@u-psud.fr}
    \affiliation{Laboratoire de Physique des Solides, CNRS, Universit\'{e} Paris-Sud, Universit\'{e} Paris-Saclay, 91405 Orsay Cedex, France}
\author{Jan Vogel}
    \email{jan.vogel@neel.cnrs.fr}
    \affiliation{Institut N\'{e}el, CNRS, 25 avenue des Martyrs, B.P. 166, 38042 Grenoble Cedex 9, France}

\begin{abstract}
Magnetic domain walls are objects whose dynamics is inseparably connected to their structure.
In this work we investigate magnetic bilayers, which are engineered
such that a coupled pair of domain walls, one in each layer, is stabilized by a cooperation of Dzyaloshinskii-Moriya interaction and flux-closing mechanism. The dipolar field mediating the interaction between the two domain walls, links not only their position but also their structure. We show that this link has a direct impact on their magnetic field induced dynamics.
We demonstrate that in such a system the coupling leads to an increased domain wall velocity with respect to single domain walls. Since the domain wall dynamics is observed in a precessional regime, the dynamics involves the synchronization between the two walls, to preserve the flux closure during motion. Properties of these coupled oscillating walls can be tuned by an additional in-plane magnetic field enabling a rich variety of states, from perfect synchronization to complete detuning.
\end{abstract}

\maketitle

\date{\today}

When several similar oscillators are coupled by a weak force, they can adjust their rhythms thanks to entrainment, similarly to the original observations of pendulum clocks by C. Huygens. The field of synchronization covers a vast amount of phenomena in daily life, nature, music, communication or engineering \cite{pikovsky2003synchronization}. In spintronics, spin-torque nano-oscillators have demonstrated great ability to synchronization via coupling through electrical current, spin-waves or dipolar field \cite{mancoff2005phase,kaka2005mutual, grollier2006synchronization, pufall2006electrical} leading to a narrower linewidth and an increased emitted power. Here we use the physics of synchronization to enhance the velocities of magnetic domain walls (DWs) in thin magnetic bilayers. The interaction between two oscillators, represented by a pair of chiral DWs, is mediated by a dipolar field which not only links the wall position  but also locks their internal structure. The present work represents a first experimental realization of a coupled system whose motion can benefit from synchronization \cite{o2017oscillators}.

\begin{figure*}[ht!]
\includegraphics[width=\textwidth]{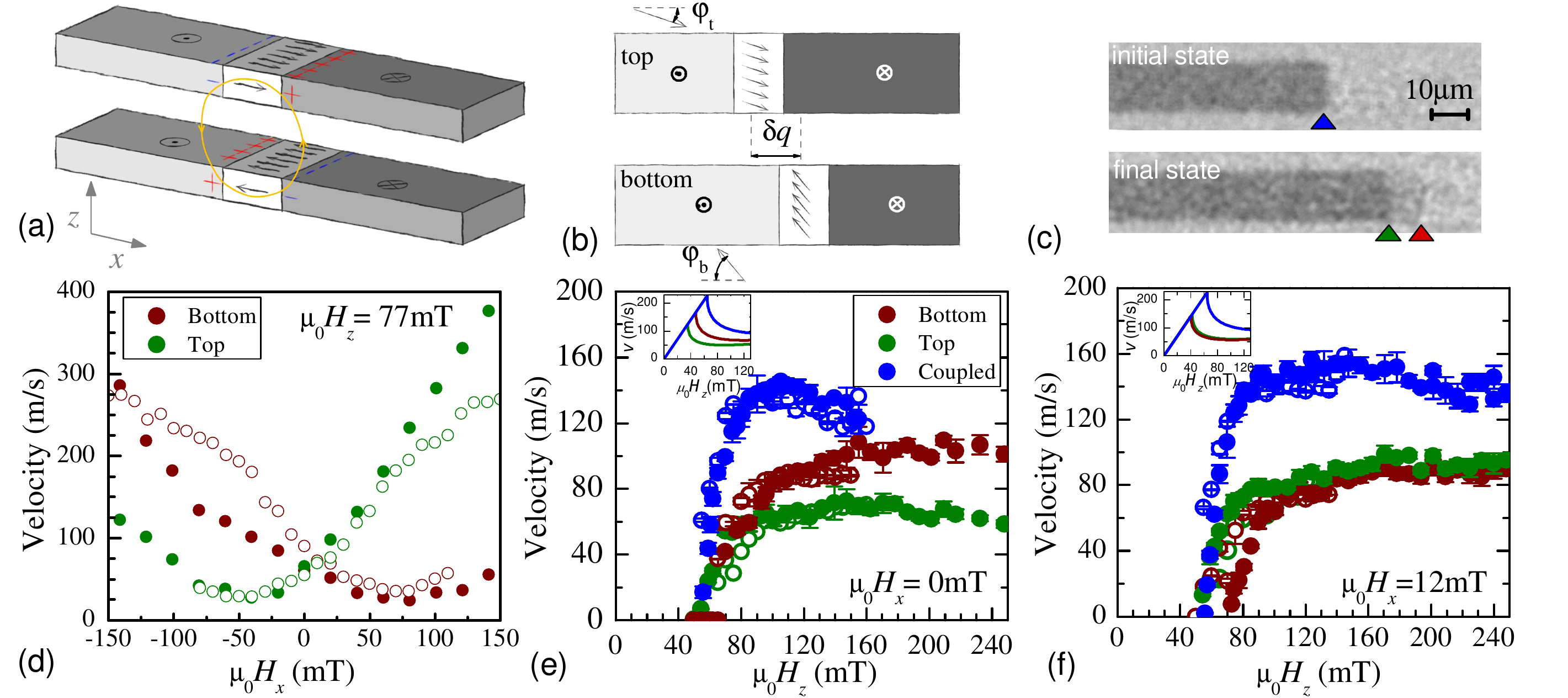}
\caption{Coupled domain walls dynamics.
(a) Magnetic bilayer track containing two up-down chiral N\'{e}el walls coupled by dipolar field. (b) When an out-of plane field $B_z$ is applied,
the top and bottom DW magnetization angles (seen separately in top view) tilt.
DWs are described by magnetization angles $\varphi_\mathrm{b}$ and $\varphi_\mathrm{t}$ inside the bottom and top walls and by their relative position $\delta q$. (c) Kerr micrographs in the initial state [coupled DW (blue triangle)] and after application of seven 30-ns, long 150-mT field pulses (bottom), which result in decoupled DWs (green=top layer, red=bottom layer). Dark, light-gray and gray contrasts inside the stripe correspond to $\uparrow\uparrow$, $\uparrow\downarrow$ and $\downarrow\downarrow$ magnetization directions, respectively.
(d) Single DW velocity at $\mu_0H_z=77$~mT for an up-down DW, as a function of an in-plane field.
(e),(f) DW velocity as a function of an out-of-plane field at $\mu_0H_x=0$~mT (e) and $+12$~mT (f). In (d-f), the color code  is the same; full and empty symbols correspond to the experiments and 2D micromagnetic simulations respectively. Insets in (e),(f) show the calculated curves by the minimal model \cite{slonczewski1972dynamics}, using the coupled DWs effective DMI constant (strong coupling approximation).
\label{Fig_1}}
\end{figure*}

Magnetic domain wall (DW) dynamics usually focuses on velocity response to a driving force (magnetic field or electrical current). However, a DW is not a simple interface described only by its position $q$, as dynamics is directly connected to its internal structure. In a minimal model \cite{slonczewski1972dynamics}, $q$ is connected to the angle $\varphi$, which describes the DW internal magnetization orientation, and they form a set of conjugated coordinates (Fig.~\ref{Fig_1}). At small driving forces, the motion is steady-state and $\varphi$ adopts a finite value, but above a threshold called Walker breakdown, the DW is no longer able to hold its structure: $\varphi$ precesses and the velocity drops down \cite{slonczewski1972dynamics, hayashi2007direct}. In such a regime, the DW behaves as an oscillator as well as a moving interface with a direct link between them. The boundary between the two regimes, set by the DW internal energy, can be efficiently controlled in ultrathin magnetic films by inducing chirality through the Dzyaloshinskii--Moriya interaction (DMI) \cite{heide2008dzyaloshinskii,Thiaville_DMI,ryu2013chiral,emori2013current}. This unavoidably also changes the DW oscillation properties such as the precession frequency.

DWs in magnetic multilayers can be coupled via direct exchange across a spacer layer \cite{yang2015domain, lepadatu2017synthetic, metaxas2010dynamic} or through dipolar fields \cite{bellec2010domain,purnama2015coupled,hrabec2017current}. The benefits of the oscillatory motion have been overlooked so far since the dynamics of these bound DWs was probed in a steady-state regime. In this Letter, we show that a pair of dipolarly coupled DWs moving above the Walker breakdown can synchronize their precession, resulting in a velocity increase. While the DWs are moved by a perpendicular magnetic field, an additional in-plane field allows tuning the oscillator properties to optimize the synchronization and maximize the velocity. This allows exploring a rich variety of synchronized states, from entrainment to complete detuning.

We study a magnetic multilayer~\cite{note_SI} of Pt(5)$\backslash$Co(1.1)$\backslash$Au(3)$\backslash$Co(1.1)$\backslash$Pt(5) (thicknesses in nanometer) with perpendicular magnetic anisotropy where the Au spacer thickness is adjusted in order to ensure purely dipolar coupling between the magnetic layers \cite{grolier1993unambiguous,hrabec2017current}. Such a structure offers a way to constructively combine two chiral interactions \cite{hrabec2017current}: the DMI, whose sign is set by the interface nature \cite{belabbes2016hund}, and the dipolar fields \cite{bellec2010domain}, whose sense is not adjustable. Through the DMI, the Pt$\backslash$Co interface serves as a source of left- and right-handed chirality \cite{yang2015anatomy, corredor2017sempa} in the DWs in the bottom and top Co layers respectively. Therefore, the  DWs have opposite chiralities and can be coupled through the stray-field in a flux closure manner illustrated in Fig.~\ref{Fig_1}(a) \cite{hrabec2017current}. To study magnetic field-induced DW motion, the films were patterned into 10~$\mu$m wide stripes using UV lithography and e-beam etching through an aluminium mask.

We first focus on the coupled DWs velocity. Polar Kerr magneto-optical microscopy was used to image the DW motion along the wires in response to sequences of 30-ns magnetic field pulses, obtained using a microcoil embedded in a silicon substrate \cite{pham2016very}. Coupled  DWs were obtained by domain nucleation on natural sample defects, after the application of a 30-ns, long 120-mT field pulse.
Different shades of magnetic contrast reveal whether coupled or single DWs are present, as shown in Fig.~\ref{Fig_1}(c) prior and after application of 150-mT pulses leading to DW decoupling. This allows investigating the individual DW behavior and thus the properties of each magnetic layer. Fig.~\ref{Fig_1}(e) reveals that the coupled DWs velocity increases up to $\approx140$~m/s at 100~mT and then gradually decreases until $\approx150$~mT where the walls decouple and  no longer travel together. The single DW velocity curves are globally similar to that of coupled DWs, but the maximum velocity is significantly lower. Additionally, the DW velocity versus an in-plane field aligned along the wires and driven by $\mu_0H_z=77$~mT was measured  to evidence the opposite chirality of single walls in each layer [see Fig.~\ref{Fig_1}(d)] \cite{vavnatka2015velocity}. Beyond the DMI sign change, the difference between single walls velocity curves indicates that the magnetic layers are slightly different.

The velocity curves reveal three distinct regimes: a negligible velocity up to 50~mT due to sample defect-induced pinning, a fast velocity increase during a depinning transition (50--70~mT) and a regime where the velocity saturates. The steady-state regime is hindered by defects as the Walker field [see the inset of Fig.~\ref{Fig_1}(e)] is close to the depinning transition \cite{DiazPardo2017}. The velocity saturation is typical for a precessional dynamics where the oscillations are perturbed by vertical Bloch lines (VBLs) propagating along the wall \cite{malozemoff1972effect}, and is correlated with the DMI \cite{yoshimura2016soliton,pham2016very}. The larger coupled DWs saturation velocity seen in Fig.~\ref{Fig_1}(e) is a direct consequence of the chiral energy increase due to the flux-closure.

\begin{figure*}[ht!]
\includegraphics[width=\textwidth]{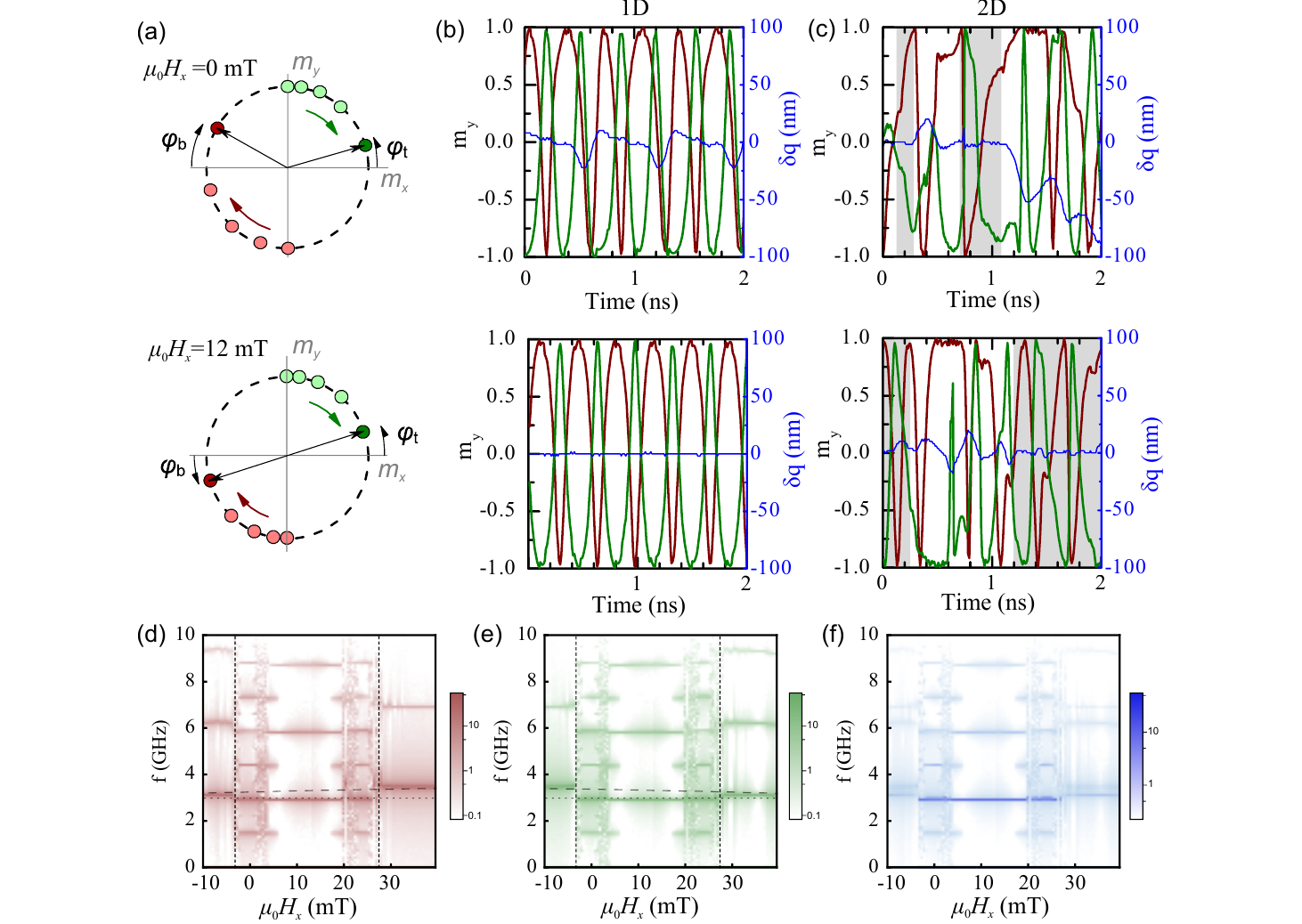}
\caption{Simulations illustrating the DW oscillation synchronization. (a) Sketch of the magnetization angle in the center of the bottom ($\varphi_\mathrm{b}$) and top ($\varphi_\mathrm{t}$) DWs. In the absence of the in-plane field (top panel) the two oscillations are detuned while the appropriate in-plane field (bottom panel) ensures their locking. (b) Time evolution of $m_y$ for a 1D wire containing a pair of up-down DWs at $\mu_0H_z=125$~mT, for $\mu_0H_x=0$~mT (top panel) and $+12$~mT (bottom panel). The coupled DWs velocity is 98~m/s [inset of Fig.~\ref{Fig_1}(e)]. Green and red curves represent the top and bottom DWs respectively. The blue curves denote the relative distance between the walls. (c) Evolution of $m_y$ at $\mu_0H_z=125$~mT for a 2D system in the middle of the stripe as indicated by the dashed line in Fig.~\ref{Fig_3}(b). (d) Power spectral density of $m_y$ calculated from 1D model [panel (b)] for a 20~ns time window, as a function of the in-plane field for the bottom (d) and top (e) DWs. Vertical dashed lines show lower and upper bounds of individual (decoupled) DW. Grey dashed and dotted lines represent the minimal model \cite{slonczewski1972dynamics} results for the coupled (using $D_\mathrm{eff}$) and single DWs (using $D_\mathrm{b}$ and $D_\mathrm{t}$) respectively. (f) Geometric mean of the values presented in panels (d) and (e) which underlines the synchronization regions. \label{Fig_2}}
\end{figure*}

Micromagnetic simulations using \textsc{MuMax} \cite{Vansteenkiste2014}, including magnetic disorder~\cite{gross2018skyrmion}, were used to determine the sample parameters (DMI coefficients $D$, disorder amplitude and damping parameter $\alpha$), by finding the best agreement with experiments [see Fig.~\ref{Fig_1}(d-f) and Supplementary materials \cite{note_SI}]. This yields $\alpha=0.3$, Co thickness fluctuations of $8\%$ and $D_\mathrm{b}=-0.73$~mJ/m$^2$ and $D_\mathrm{t}=0.50$~mJ/m$^2$ in the bottom and top layer, respectively, manifesting a change of interface quality along the growth direction. These parameters are used to calculate the coupled DWs dynamics [Fig.~\ref{Fig_1}(e)] without further adjusting any of the parameters.

The coupling between DWs is mostly mediated by the stray-field arising from the domains, and can be decomposed into two components. The stray-field acting on the domains generates an attractive spring force between the DWs. The stray-field acting on the DW magnetization modifies the DW energy via a magnetic flux-closure and has the same symmetry as DMI \cite{hrabec2017current}. In a strong coupling approximation, with negligible separation $\delta q$ and antiparallel magnetization of the two DWs, the coupled DWs can be represented by a single DW with an effective DMI $D_\mathrm{eff}=1.08$~mJ/m$^2$ (73~\% larger than the average of the DMI coefficients), estimated from static calculations \cite{note_SI}. Hence, increased Walker field and velocities shown in the inset of Fig.~\ref{Fig_1}(d-e) is in qualitative agreement with experimental data. Note that the increase scales with the layer magnetization and spacer thickness.

Since the DMI is not equivalent in the two layers, an in-plane field can be used to tune and symmetrize the system by reducing (reinforcing) the stability of the bottom (top) DW \cite{note_SI}. With $\mu_0H_x=+12$~mT applied in addition to the out-of-plane field, the isolated up-down DW velocities [Fig.~\ref{Fig_1}(f)] approach each other (Down-up DWs require opposite in-plane field sign but we focus on up-down DWs without any loss of generality). Interestingly, coupled DWs in the symmetrized system display an unchanged maximum velocity, but are more robust, since they do not decouple up to $\approx250$~mT while maintaining a constant velocity.

\begin{figure}[ht!]
\includegraphics[width=\columnwidth]{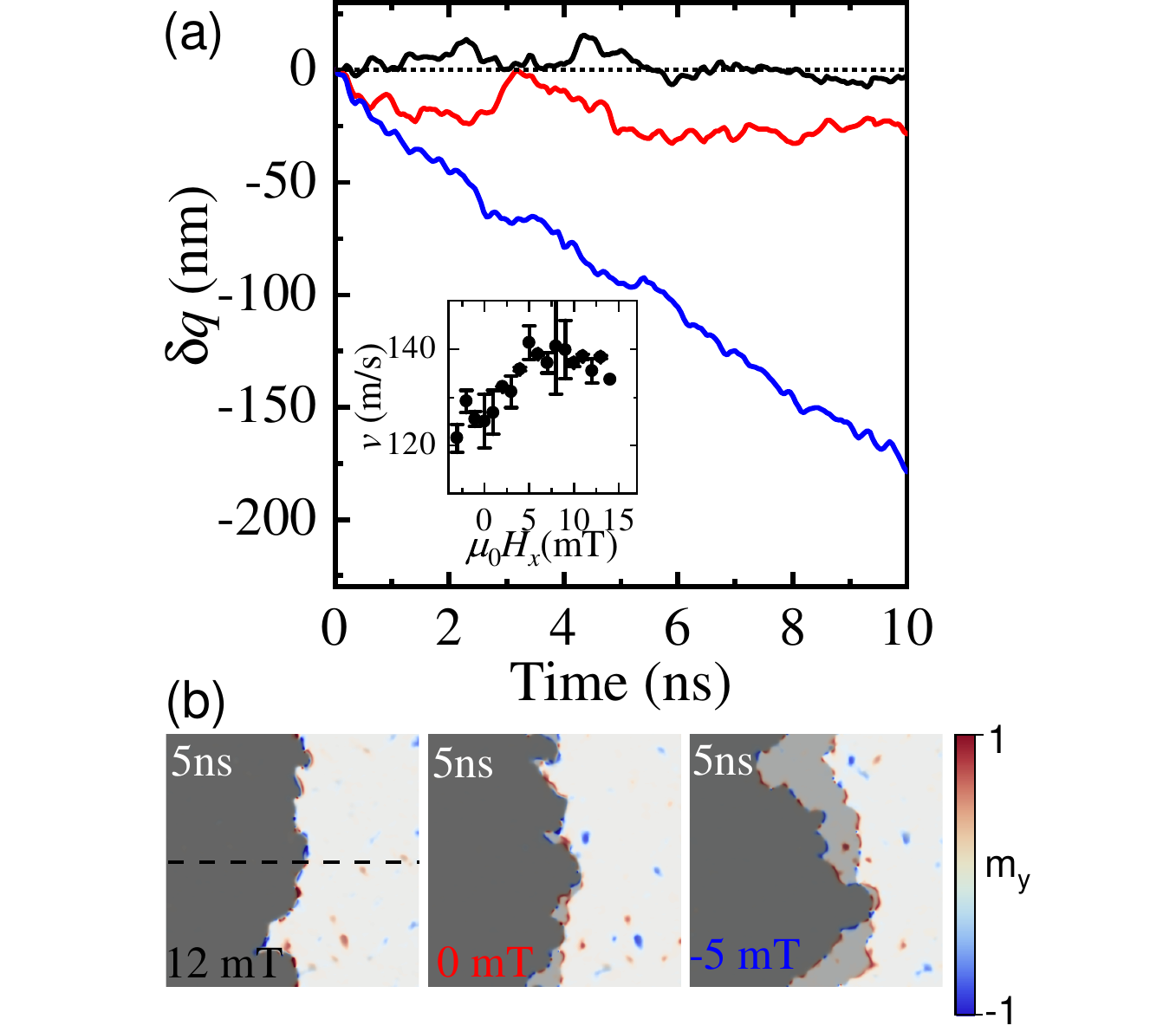}
\caption{Simulation of DW motion in a 2D disordered medium. (a) Calculated evolution of the average distance between bottom and top DWs at $\mu_0H_z=125$~mT for $\mu_0H_x=+12$ (black), $=0$~mT (red) and $-5$~mT (blue). The inset shows the resulting coupled DWs velocities as a function of the in-plane field. (b) Snapshots of the two DWs dynamics at 5~ns corresponding to panel (a). The dark gray contrast corresponds to $\uparrow\uparrow$ state, light gray to $\uparrow\downarrow$ state and white contrast to the $\downarrow\downarrow$ state. The $m_y$ component is reflected by red ($m_y=1$) and blue ($m_y=-1$) color scheme. The size of the moving box centered on the top wall is $1\times1$~$\mu$m$^2$. \label{Fig_3}}
\end{figure}

When an out-of-plane field is applied, the magnetization directions $\varphi_b$ and $\varphi_t$ inside the bottom and top DWs change proportionally to the field amplitude and, at the Walker breakdown, the DWs turn into chiral Bloch walls ($\varphi_\mathrm{b}=\varphi_\mathrm{t}=-\pi/2$). However, since experimentally the depinning field is higher than the Walker field, $\varphi_b$ and $\varphi_t$ cannot be stationary. Therefore, maintaining the velocity enhancement requires that the DW coupling remains valid even in such a situation, implying that $\varphi_\mathrm{b}=\varphi_\mathrm{t}$.

We first start with 1D modeling of the dynamics at $\mu_0H_z=125$~mT, for which the DW magnetization undergoes regular precession (Fig.~\ref{Fig_2}(a)): when the two DWs maintain $\varphi_\mathrm{b}=\varphi_\mathrm{t}$, they move \textit{together} and \textit{fast} contrary to the case when the DWs do not lock and lose the benefits of the flux-closure. The dynamics is illustrated by 1D micromagnetic simulations~\cite{note_1Dsim} [Fig.~\ref{Fig_2}(b)]. Under a 12~mT in-plane field, the two DWs have similar precession frequencies and can lock-in, since the DMI difference  is compensated by the in-plane field \cite{note_SI}, with $\delta q\approx0$. As a consequence, the coupled DWs velocity is larger than the single DWs (98~m/s as compared to 57~m/s). For a detuned system [see Fig.~\ref{Fig_2}(b) with no applied field], the single DW precession frequency difference makes synchronization more complex: the relationship $\varphi_\mathrm{b}=\varphi_\mathrm{t}$ is not perfectly fulfilled and results in $\delta q$ oscillations. For even larger detuning, synchronization becomes impossible and the walls separate.

To explore synchronization in more detail, we have computed $m_y$ spectral densities as a function of the in-plane field [Figs.~\ref{Fig_2}(d) and (e)] for each DW. The DWs are only coupled in the (-3--27)-mT field interval. Outside of this interval, the DWs are spatially separated and their magnetizations oscillate at their natural frequencies estimated by the minimal model \cite{slonczewski1972dynamics} (grey dashed lines). The geometrical average of these spectral densities [Fig.~\ref{Fig_2}(f)] emphasizes the synchronization. The simplest synchronized state is found in a  (+6--+18)-mT field range, centered around the perfectly tuned situation. It displays a ground precession frequency $f_0$ of 3~GHz at $\mu_0H_z = 125$~mT, corresponding to the precession of a single DW with the effective DMI (grey dotted lines). This range is surrounded by zones of more complex synchronization with the appearance of $f_0/2$ frequencies, frequency emission over a continuous frequency range, and suggests more a chaotic relationship \cite{note_SI}.

The situation in a real medium, with 2D degrees of freedom, is more complex since the DW oscillations are perturbed by VBLs and structural disorder. In Fig.~\ref{Fig_3}(b) are displayed snapshots of the DW motion under $\mu_0H_z=125$~mT for three different in-plane fields (see also Supplementary movies  \cite{note_SI_movies}). For a $+12$-mT in-plane field, the DWs mostly overlap with small separation fluctuations [Fig.~\ref{Fig_3}(a)], which underlines the strong coupling efficiency. For a slightly detuned system (here at zero in-plane field), a larger separation is observed, together with a velocity decrease [inset of Fig.~\ref{Fig_3}(a)]. Ultimately, at even larger detuning (for $\mu_0H_x\leq-5$~mT, similarly to the 1D model), the two DWs are not able to hold together and their separation gradually increases until they decouple and move independently.

To locally probe the magnetization variation correlation, i.e. the local DW coupling, the time evolution of magnetization at the center of the DW [the linecut indicated in Fig.~\ref{Fig_3}(b)] is shown in Fig.~\ref{Fig_2}(c), for $\mu_0H_x=+12$ and 0~mT (without any structural disorder for simplicity). Even with the symmetrizing field, synchronization is not always observed but time intervals of synchronization are more frequent and longer than in the detuned system. During the non-synchronized intervals, DWs are misaligned since the coupling is lost (and vice versa). However, synchronization can be recovered in this 2D situation, since at a given time, some portions along the DWs are still coupled, as observed in the Supplementary movies \cite{note_SI_movies}. These parts drag and accelerate the rest of the DW pair and the DWs decouple once the synchronized sections are no longer able to drag the separated DWs with lower mobilities.

To summarize, we have experimentally shown that the dipolar coupling of magnetic domain walls in superposed layers leads to a large velocity increase.
This phenomenon is robust against domain wall magnetization precession and disorder.
At first sight, especially when DWs have different properties, this is unexpected and implies that the wall magnetization dynamics synchronizes.
Systems where the motion of coupled oscillators can benefit from their synchronization have
been recently theoretically described as "swarmalators" \cite{o2017oscillators}.
The study here can be seen as an experimental realization of such a system, with two swarming objects.

\begin{acknowledgments}

We thank to Joo-Von Kim, Julie Grollier for fruitful discussions and to Jacques Miltat for critical comments on the manuscript. This work has been supported by the Agence Nationale de la Recherche (France) under Contract No. ANR-14-CE26-0012 (Ultrasky), ANR-17-CE24-0025 (TopSky), ANR-09-NANO-002 (Hyfont), the RTRA Triangle de la Physique (Multivap).
\end{acknowledgments}

\section{Supplementary Information~1: Sample details and micromagnetic parameters determination}

Samples were grown by ultra-high\ vacuum evaporation with a base pressure of $10^{-10}$~mBar. The multilayers of Pt(5)$\backslash$Co(1.1)$\backslash$Au(3)$\backslash$Co(1.1)$\backslash$Pt(5) were deposited on a high-resistive silicon with a native oxide layer (all thicknesses in nanometers). The magnetization $M_\mathrm{s}=1.42$~MA/m and uniaxial anisotropy $K=1.71$~MJ/m$^3$ were determined by SQUID measurements. DMI, damping and sample disorder have been determined by comparing experiments to micromagnetic simulations, carried out using the MuMax3 code \cite{Vansteenkiste2014}, in  $1\times1$~$\mu$m$^2$ stripes, with periodic boundary conditions along the direction orthogonal to the DW motion, and with  2~nm $\times$ 2~nm $\times$ 1.1~nm cell size .

\subsection{DMI}

\begin{figure}[h!]
\includegraphics[width=\columnwidth]{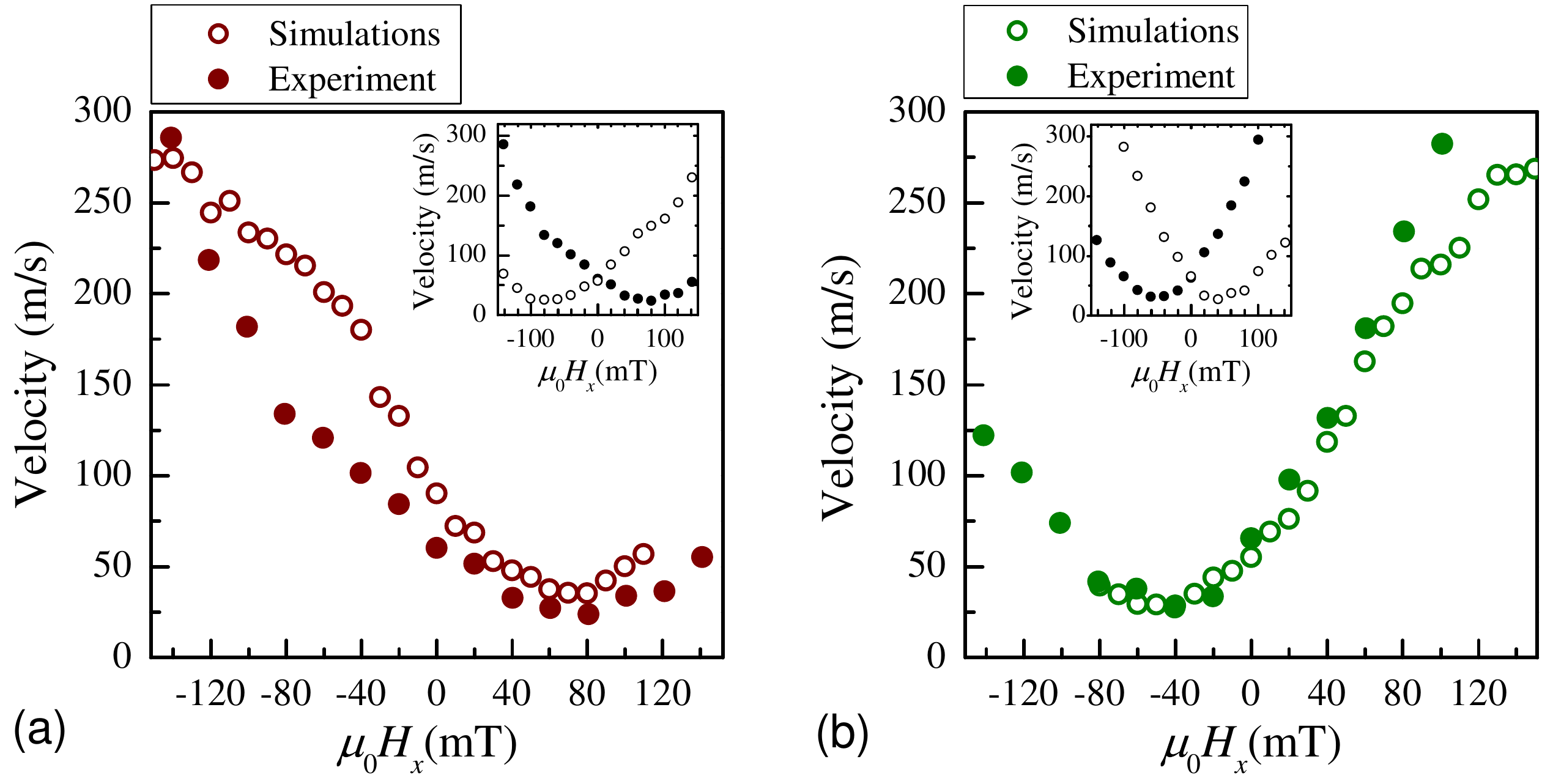}
\caption{\textbf{DMI measurements.} DW velocities for the case of an up-down DW in the bottom (a) and top (b) layers as a function of an-in plane field $\mu_0H_x$. The amplitude of the out-of plane field is fixed to $\mu_0H_z=77$~mT. Insets represent experimental data for up-down (full symbols) and down-up (empty symbols) DWs.}  \label{FigSup_1}
\end{figure}

In order to measure the DMI in the individual layers, we have used the in-plane magnetic field-based method. Here the in-plane magnetic field modifies the nature and the energy of the DW. The DW energy
acquires a maximum value when the applied in-plane field is equal and opposite to the stabilising DMI field i.e. when the DW acquires a Bloch form. Therefore a measurement of the minimum of velocity provides a direct measure of the DMI. To avoid any problems with the DW propagation in the creep regime\cite{vavnatka2015velocity,pellegren2017dispersive}, we have applied a magnetic field of $\mu_0H_z=77$~mT close to the flow regime. We have checked that any parasitic out-of-plane field arising from the misalignment of the in-plane field was eliminated by measuring the DW velocities for up-down and down-up  cases presented in insets of Supplementary Fig.~\ref{FigSup_1}(a) and (b). The main experimental curves are shown in Supplementary Fig.~\ref{FigSup_1} for the case of a DW introduced in the bottom (a) and top (b) layers. The best agreement between the experimental and micromagnetic datasets was found for $D_\mathrm{b}=-0.73$~mJ/m$^2$ and $D_\mathrm{t}=+0.50$~mJ/m$^2$.

\subsection{Damping $\alpha$}

Given the values of DMI in the bottom and top layers, velocities of isolated DWs in the bottom and top layers were numerically calculated and are presented in Supplementary Fig.~\ref{FigSup_alpha}. The best match between the experimental data and the micromagnetic simulations is found for $\alpha=0.3$.

\begin{figure}[h!]
\includegraphics[width=\columnwidth]{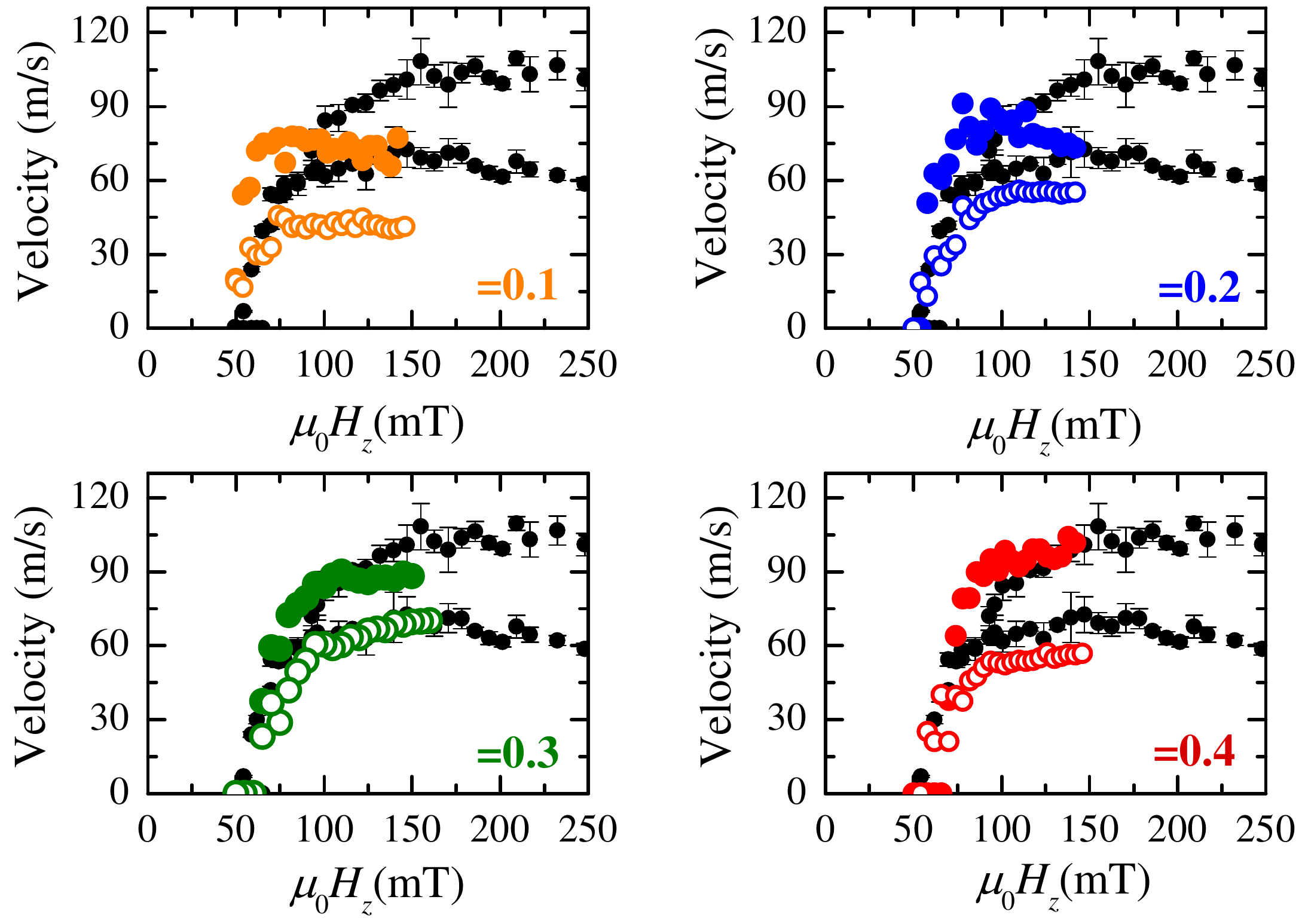}
\caption{\textbf{Effect of damping $\alpha$.} Calculated DW velocities for an isolated DW in the top and bottom layers respectively for various parameters $\alpha$.}  \label{FigSup_alpha}
\end{figure}

\subsection{Disorder}

Disorder is included by a random fluctuation following a normal distribution of the ferromagnetic layers thickness $t$ between columnar grains arranged in a Voronoi fashion \cite{gross2018skyrmion}. The average lateral grain size is fixed to 15~nm. Since the micromagnetic code requires a computational cell with a constant thickness over the whole sample, the saturation magnetization is varied from grain to grain as $M_\mathrm{s}t/t_0$. Averaged over the thickness, the uniaxial anisotropy $K_\mathrm{u}$ and the DMI constants $D$ are also directly modified in each grain, i.e. $K_\mathrm{u} = K_\mathrm{s}/t$ and $D= D_\mathrm{s}/t$. Effect of the disorder is shown in Supplementary Fig.~\ref{FigSup_disorder}(a) revealing how the disorder cuts off the low-field regime. A secondary effect of the disorder is a spontaneous nucleation of reversed domains at high magnetic fields which sets the upper bound of used magnetic fields in our simulations.

\begin{figure}[h!]
\includegraphics[width=\columnwidth]{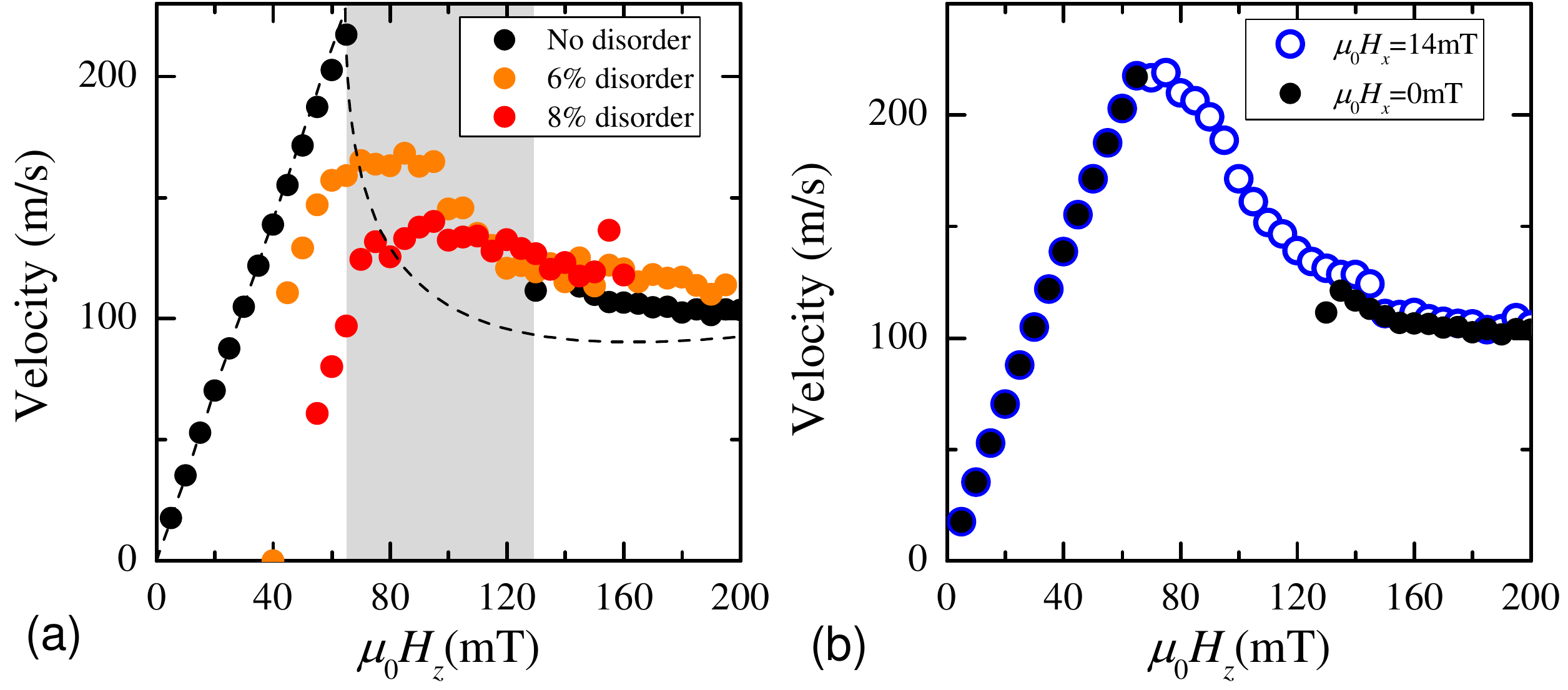}
\caption{\textbf{Effect of disorder on coupled DW dynamics.} (a) Calculated coupled DW velocities in the case of various disorder. The grey range corresponds to the case when the DWs are decoupled in the regime with no disorder. The dashed line corresponds to the analytically calculated 1D model case. (b) Coupling of the DWs can be restored with a symmetrizing in-plane field $\mu_0H_x=+14$~mT.}  \label{FigSup_disorder}
\end{figure}

\subsection{Micromagnetic parameters}

The micromagnetic parameters obtained by the above-described 2D micromagnetic calculations (including disorder) fitting the DW dynamics which are presented in Supplementary Table~\ref{Table1}.

\begin{table}[h!]\begin{tabular}{ll|ccc}
  \hline\hline
  Name & Label & \multicolumn{2}{c}{Value} & Unit \\
       &       & Top & Bottom              & \\
      \hline\hline
  Thickness                      & $t$            & \multicolumn{2}{c}{1.1}                     & nm \\
  Exchange constant              & $A$            & \multicolumn{2}{c}{16}                      & pJ/m\\
  Magnetization                  & $M_\mathrm{s}$ & \multicolumn{2}{c}{1.45}                    & MA/m\\
  Anisotropy                     & $K$            & \multicolumn{2}{c}{1.71}                    & MJ/m$^3$\\
  Dzyaloshinskii-Moriya constant & $D$            & 0.50 & -0.73                                & mJ/m$^2$\\
  Damping coefficient            & $\alpha$       & \multicolumn{2}{c}{0.3}                     & \\
  Gyromagnetic factor            & $\gamma$       & \multicolumn{2}{c}{$1.909 \times 10^{11}$}  & rad s$^{-1}$ T$^{-1}$\\
  Grain size                     &                & \multicolumn{2}{c}{15}                      & nm\\
  Thickness fluctuations         &                & \multicolumn{2}{c}{8}                       & \%  \\
  \hline\hline
 \end{tabular}
 \caption{\textbf{Material parameters used for the micromagnetic simulations.}}
 \label{Table1}
 \end{table}


\section{Supplementary Information~2: Domain wall energy and dynamics}

\subsection{Domain wall energy}

The DW dynamics is connected to the DW energy. Therefore, to understand the effect of DW coupling, we explicitly derive the different terms that form the DW energy $\sigma$. For a single wall, it reads as
\begin{equation}
\sigma = \sigma_0 -\pi D \cos\varphi + \delta_\mathrm{N} \cos^2\varphi
\end{equation}
where $\varphi$ represent the internal magnetization orientation (conventionally $\varphi = 0$ represents the N\'eel wall favored by a positive $D$) and where the first term, $\sigma_0$, corresponds to the Bloch wall energy, related to the Heisenberg exchange and effective anisotropy energies, the second term corresponds to the DMI energy and the last term corresponds to the dipolar energy cost related to the N\'eel wall configuration. This last term originates from the volume charges created by the N\'eel wall and can be expressed as $\delta_\mathrm{N}\approx\mu_0M_\mathrm{s}^2t\ln2/\pi$ [\onlinecite{Thiaville_DMI}], with $t$ the film thickness and $\Delta$ the domain wall width.

When two walls are dipolarly coupled in a symmetric fashion \cite{hrabec2017current}, walls have opposite chiralities, which satisfies both DMI and magnetic flux closure at equilibrium. Therefore in a strong coupling limit (i.e. negligible separation delta $q$ and $\varphi_\mathrm{b}=\varphi_\mathrm{t}=\varphi$) the wall energy reads \cite{hrabec2017current}:
\begin{equation}\label{equation_energies}
\sigma = \sigma_0 -\pi \langle D\rangle \cos\varphi + \delta_\mathrm{N} \cos^2\varphi - \delta_\mathrm{W-W}\cos^2\varphi -\delta_\mathrm{D-W}\cos\varphi
\end{equation}
where the angle $\varphi$ is defined in Fig.~1(b) and $\langle D\rangle = \frac12(|D_\mathrm{b}|+|D_\mathrm{t}|)$. Two new terms appear. $\delta_\mathrm{W-W}$ corresponds to the dipolar coupling between the N\'eel charges of both DWs, and has therefore the same physical origin as the dipolar cost of the N\'eel walls. For a spacer thickness lower than the DW width $\Delta$, their absolute magnitudes are expected to be close and therefore $\delta_\mathrm{N}+\delta_\mathrm{W-W}$ can be neglected. $\delta_\mathrm{D-W}$ corresponds to the coupling of the wall magnetization with the magnetic field created by the domains (flux closure mechanism) and gives an additional chiral energy.

\begin{figure}[h]
  \begin{center}
    \includegraphics[width=0.8\columnwidth]{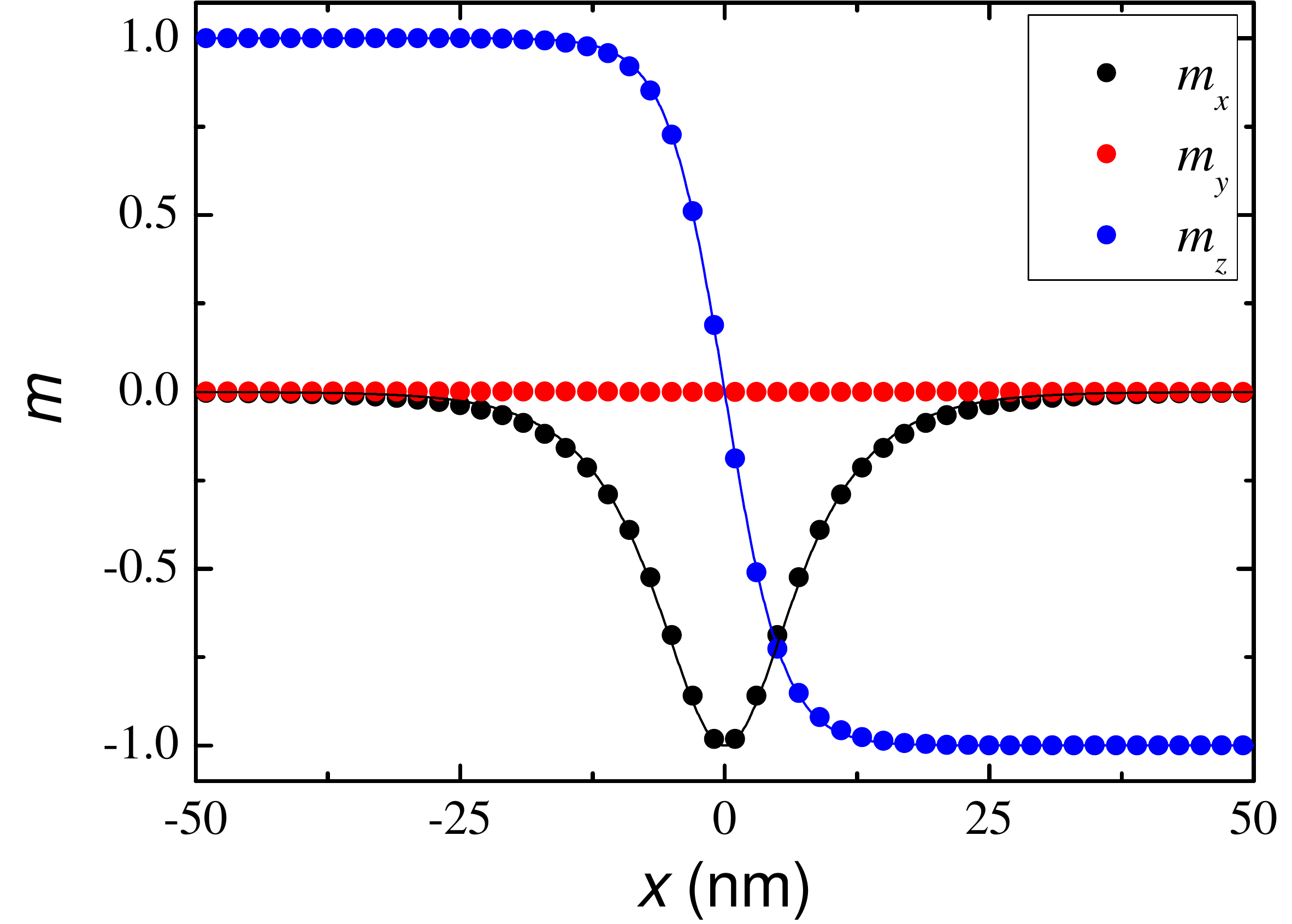}
  \end{center}
  \caption{\label{fig:DW_Profile} DW profile in a bilayer (similar profile is obtained on a single DW).}
\end{figure}

To evaluate the different terms, only the DW width $\Delta$ and $\delta_{\mathrm{D-W}}$ have to be determined from micromagnetic calculations. Isolated and coupled DWs have been relaxed (see Figure~\ref{fig:DW_Profile})
and show a close DW width ($\Delta = 5.54$ and 5.50~nm respectively for coupled and isolated DWs, determined from the magnetization profile using the Thiele wall width definition $\Delta^{-1}=\frac12\int(\partial \mathbf m/\partial x)^2\mathrm{d}x$). To calculate $\delta_\mathrm{W-W}$ we have calculated coupled DWs with $\varphi = 0$
(...$\frac{\uparrow\uparrow\uparrow\rightarrow \downarrow\downarrow\downarrow}{\uparrow\uparrow\uparrow\leftarrow\downarrow\downarrow\downarrow}$...) and $\varphi = \pi$
(...$\frac{\uparrow\uparrow\uparrow\leftarrow\downarrow\downarrow\downarrow}{\uparrow\uparrow\uparrow\rightarrow \downarrow\downarrow\downarrow}$...).
The average between the two  energies corresponds to the non chiral energy ($\sigma_{\mathrm{nc}}\approx\sigma_0$) and the half difference to the chiral energy ($\sigma_{\mathrm{c}} = \pi \langle D\rangle + \delta_\mathrm{D-W}$). We find $\sigma = 10.2$~mJ/m$^2$ and $17.0$~mJ/m$^2$ for both configurations and therefore $\sigma_{\mathrm{nc}} = 13.6$ and $\sigma_\mathrm{c}=3.4$~mJ/m$^2$. Subtracting the DMI contribution to the chiral energy leads to $\delta_\mathrm{D-W} = 1.45$~mJ/m$^2$. It may be convenient to convert the chiral energy to an effective DMI constant $D_\mathrm{eff}= \sigma_\mathrm{c}/\pi=1.08$~mJ/m$^2$ and represent the coupled wall energy as
\begin{equation}
\sigma = \sigma_0 -\pi D_\mathrm{eff} \cos\varphi.
\end{equation}

\subsection{Domain wall dynamics}

In the minimal one-dimensional model\cite{slonczewski1972dynamics}, the DW dynamics is described by the DW mobility in the steady state regime $\gamma\Delta/\alpha$ and the Walker field $H_\mathrm{w}$. While the mobility of isolated and coupled DW is similar due to the similar DW width, the Walker field strongly depends on the situation, due to the different DW energies. The Walker field is expressed as

\begin{equation}
 H_\mathrm{W} = \alpha\sin\varphi_\mathrm{W}\left(\frac{\pi}{2}|H_D|-H_K\cos\varphi_\mathrm{W}\right)
\end{equation}
with $\cos\varphi_\mathrm{W} = \frac14(\delta-\sqrt{\delta^2+8})$, $\delta = \pi|H_D|/2H_K$ and $H_D=D/M_\mathrm{s}\Delta$ and $H_K = 2K_{\mathrm{DW}}/M_\mathrm{s}$. $H_D$ is related to the DMI (real or effective) and $H_K$ is related to the dipolar induced DW anisotropy. This last term strongly depends on the situation, as for isolated DWs, $K_{\mathrm{DW}} = \delta_\mathrm{N}/2\Delta$ and for coupled DWs, $K_{\mathrm{DW}} = 0$ due to the compensation between $\delta_\mathrm{N}$ and $\delta_{\mathrm{W-W}}$. From our parameters, we deduce a Walker field of 48, 35 and 65~mT respectively for isolated DW in the bottom layer, isolated DW in the top layer, and coupled DWs. Note that the increase of the Walker field in the coupled layer is due to the increase of the chiral DW energy arising from the flux closure mechanism. The difference between the Walker field in the bottom and top layer is due to the difference in DMI constants.

Experimentally and in simulations, we have shown that applying a 12~mT field along the normal of the DW can tune the dynamics and make the system more symmetric. This value can be justified with the present calculations. Applying an in-plane field, the Walker field becomes
\begin{equation}
 H_\mathrm{W} = \alpha\sin\varphi_\mathrm{W}\left[\frac{\pi}{2}\left(|H_D|\pm H_x\right)-H_K\cos\varphi_\mathrm{W}\right]
\end{equation}
and where $\delta = (\pi/2)(|H_D|\pm H_x)/H_K$ and the $\pm$ sign accounts for the sign of $H_D$. Neglecting the variation of $\Delta$ gives a very good approximation to determine the $H_x^\mathrm{sym}$ field that provides similar Walker fields as $H_x^\mathrm{sym} = -\langle H_\mathrm{DMI}\rangle /2$. We then obtain $\mu_0H_x^\mathrm{sym} = 14.2$~mT. Without the approximation of constant $\Delta$, one gets $\mu_0H_x^\mathrm{sym} = 14.3$~mT. The slight disagreement between those simple calculations and the experiments or simulations is due to the fact that we have neglected the variation of the DW width with the applied in-plane field, which slightly changes the values of $H_D$. The comparison between one-dimensional model and simulations are presented in Supplementary Fig.~\ref{fig:DW_dyna}.
\begin{figure}[h]
  \begin{center}
    \includegraphics[width=\columnwidth]{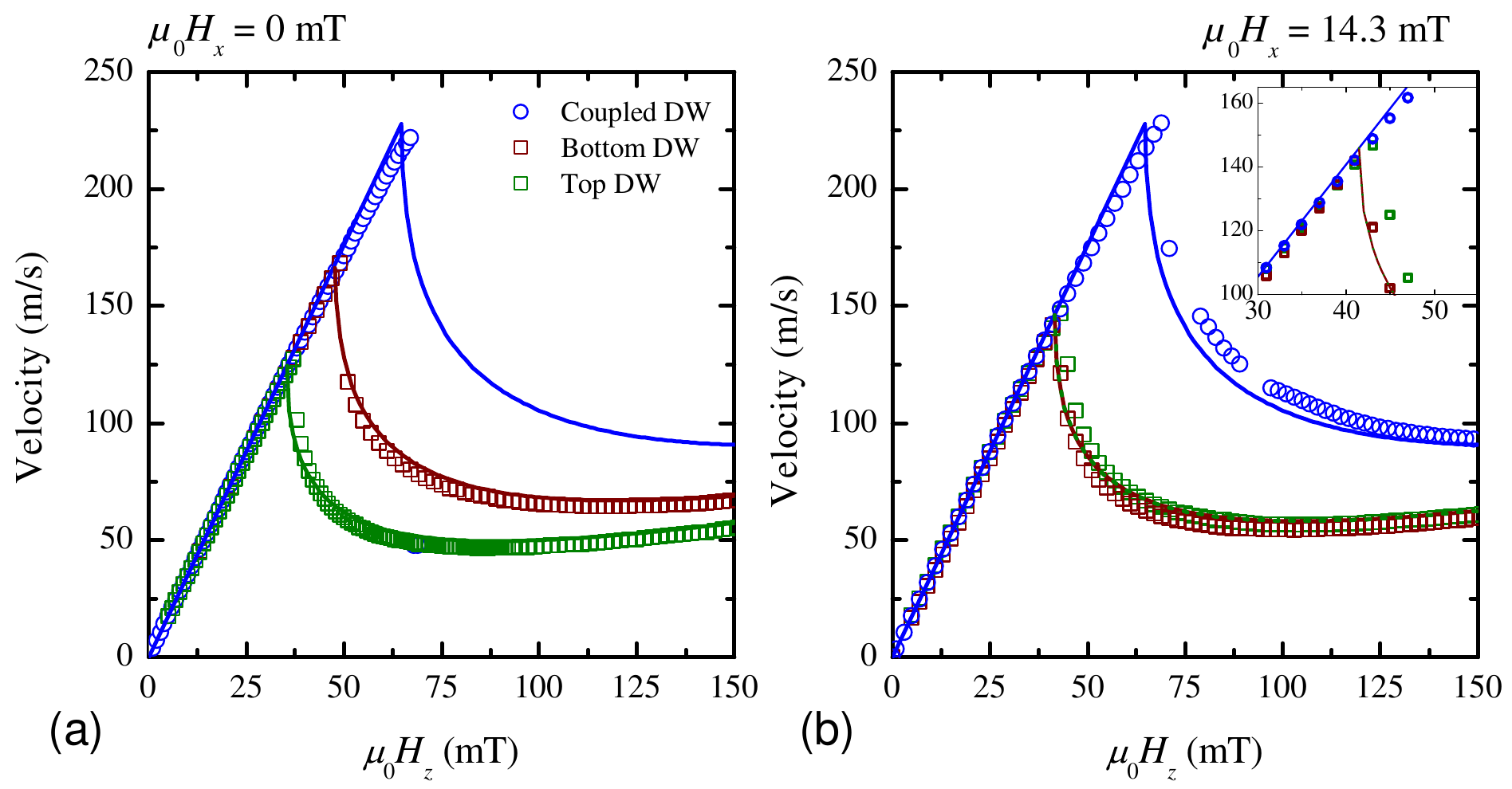}
  \end{center}
  \caption{\label{fig:DW_dyna} \textbf{Comparision between analytical calculations and 1D simulations.} Comparison between minimal analytical model (lines) and 1D simulations (dots), for an applied in-plane field of (a) 0 and (b) 14.3~mT. Note that at 14.3~mT, the simulations show that isolated DWs in each layers are not exactly superimposed contrary to the model, which was expected as the compensation field in the simulations is about 12~mT. In the coupled DW dynamics simulations, the missing points correspond to simulations where the DWs are decoupled, due to insufficient coupling.}
\end{figure}
%


\section{Supplementary Information~3: Synchronization}

\subsection{Domain wall oscillations in single layer}

\begin{figure}[h]
  \begin{center}
    \includegraphics[width=\columnwidth]{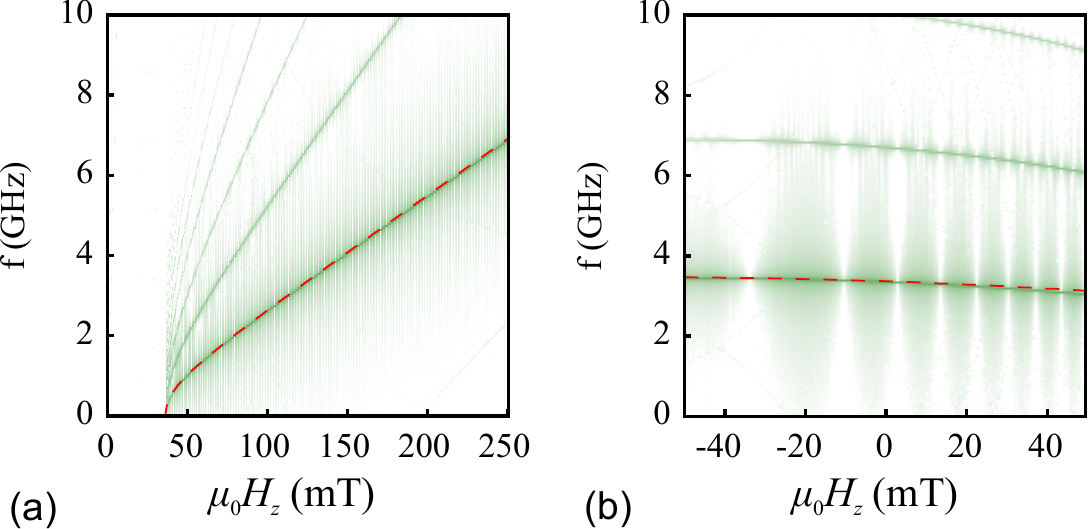}
  \end{center}
  \caption{\label{SingleLayerFreq} (a) Power spectral density of a single DW motion calculated in a 1D stripe with a parameters of the top DW as a function of an out-of-plane field at $\mu_0H_x=0$~mT. The dashed red line corresponds to the $q-\varphi$ model calculations. (b) Power spectral density of a single DW motion calculated in a 1D stripe with a parameters of the top DW as a function of an in-plane field at $\mu_0H_z=125$~mT. }
\end{figure}

To verify the validity of the Slonczewski $q-\varphi$ model, we have calculated power spectral densities for a single layer case (with micromagnetic parameters of the top magnetic layer). Supplementary Figs.~\ref{SingleLayerFreq}(a) and (b) show that the analytical model (red curves) and micromagnetic calculations are in good agreement.

\subsection{Complex synchronized states}

\begin{figure*}[h]
  \begin{center}
    \includegraphics[width=18cm]{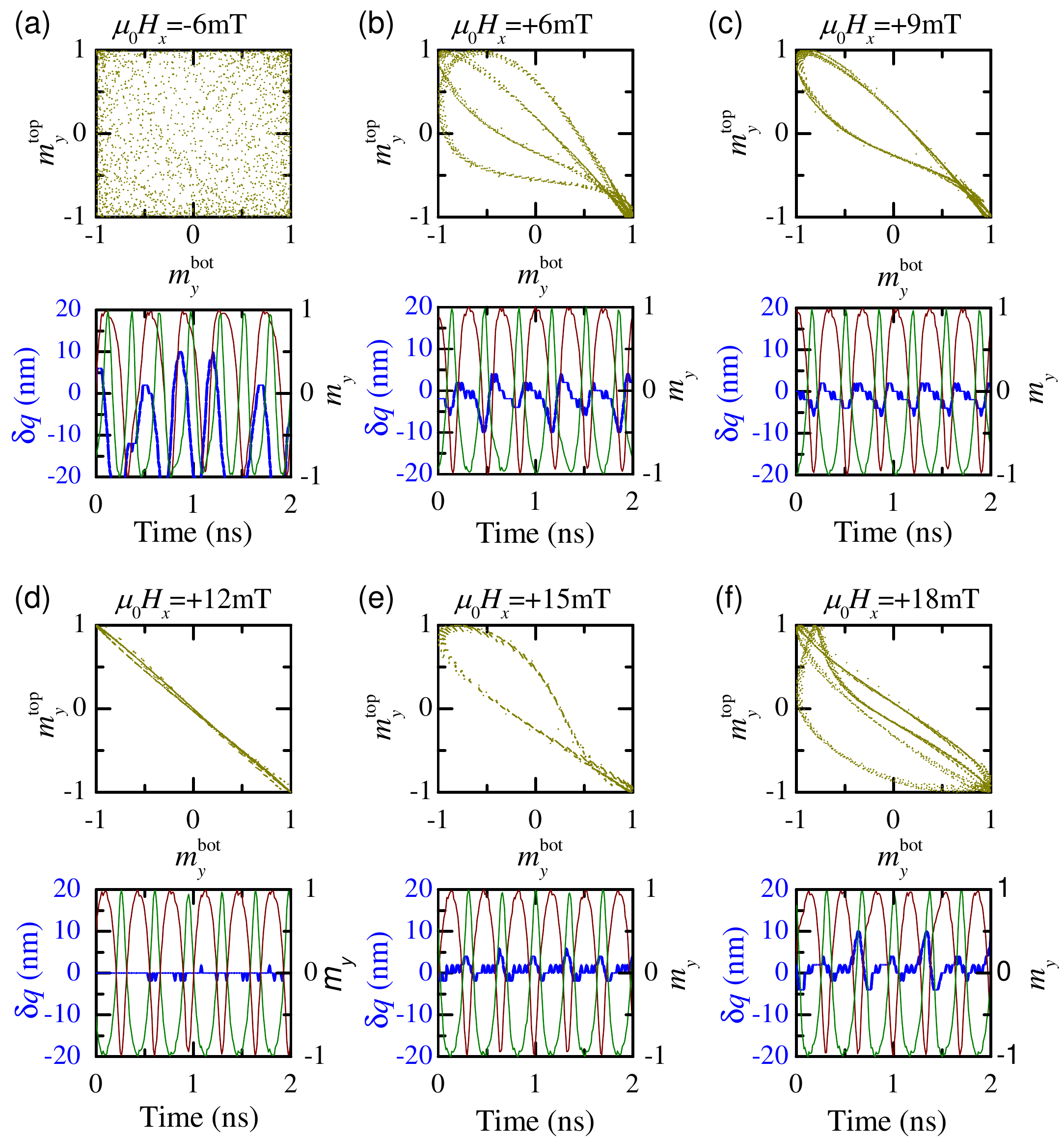}
  \end{center}
  \caption{\label{fig_Sup_Lys} Lissajous curves corresponding to the data presented in Fig.~2(b),(d) and (e) (i.e. for 1D micromagnetic calculations) for various in-plane magnetic fields where $\mu_0H_z=125$~mT. Bottom panel indicates mutual distance between DWs $\delta q$ and $m_y$ component in the center of the DW (red - bottom layer, green - top layer).}
\end{figure*}

We have supposed in Fig.~1(b) that the synchronization takes place only via the $\varphi_\mathrm{b}=\varphi_\mathrm{t}$ rotation. However, since the DWs are not necessarily perfectly aligned, the energy landscape can allow synchronous magnetization rotation via different cases. Supplementary Fig.~\ref{fig_Sup_Lys} shows Lissajous curves for the case of $\mu_0H_z=125$~mT for different in-plane fields. Such curves here illustrate mutual phase shift of $m_y$ component. Fig.~\ref{fig_Sup_Lys}(a) reveals that in the case of $\mu_0H_z=-6$~mT the two DWs are completely decorrelated as expected from Fig.~2. On the other hand, the case of $\mu_0H_z=+12$~mT shown in Supplementary Fig.~\ref{fig_Sup_Lys}(d) represents the case where the magnetization rotation takes place via the $\varphi_\mathrm{b}\approx\varphi_\mathrm{t}$ and $\delta q\approx0$ case. However, Supplementary Figs.~\ref{fig_Sup_Lys}(b,c,e,f) show that the synchronous reversal does not necessarily require $\delta q=0$ and the oscillatory variations of $\delta q$ can result in rich variety Lissajous curves.

Similar micromagnetic calculations were performed for the case presented in Fig.~2(c), i.e. for 2D micromagnetic calculations without structural disorder where the presence of VBLs is allowed. Lissajous curves for the case of  $\mu_0H_x=+12$~mT, $\mu_0H_x=+9$~mT and $\mu_0H_x=+12$~mT in the presence of OOP field $\mu_0H_z=125$~mT are presented in Supplementary Fig.~\ref{fig_Sup_DisorSync}. The structure of the DW is deduced in the middle of the strip width as indicated in Fig.~3(b). Since the two DWs do not necessarily travel together all the time but only in certain time intervals, the data shown in top panel of  Supplementary Fig.~\ref{fig_Sup_DisorSync} is polluted by the cases where the DWs are separated. The synchronization patterns similar to those presented in Supplementary Fig.~\ref{fig_Sup_Lys} become visible once we take into account only cases when the two DWs are close (here $\delta q <6$~nm for instance).


\begin{figure*}[h]
  \begin{center}
    \includegraphics[width=18cm]{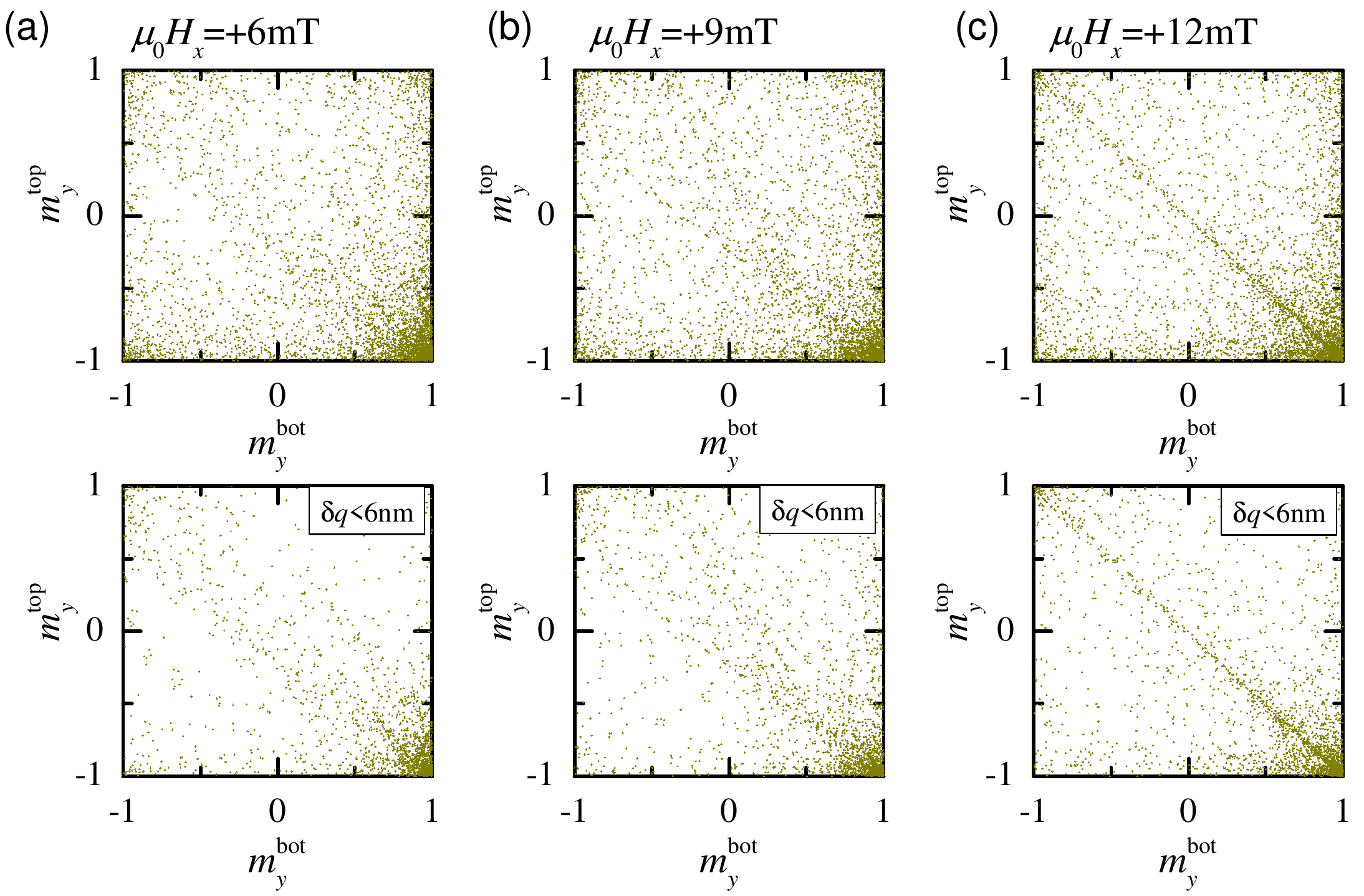}
  \end{center}
  \caption{\label{fig_Sup_DisorSync} Lissajous curves corresponding to the data presented in Fig.~2(c) (i.e. for 2D micromagnetic calculations) for various in-plane magnetic fields where $\mu_0H_z=125$~mT. Top panel displays all the data while bottom panel shows only filtered data for the case where $\delta q < 6$~nm.}
\end{figure*}

\end{document}